\documentclass[journal,twocolumn]{IEEEtran}

\usepackage{tabularx}
\usepackage{graphicx}
\usepackage{color}
\usepackage{dcolumn}
\usepackage{bm}
\usepackage[latin1]{inputenc}
\newcommand*{\rom}[1]{\expandafter\@slowromancap\romannumeral #1@}
\usepackage{epstopdf}
\usepackage{amsmath}
\usepackage{multirow}
\usepackage[normalem]{ulem}
\useunder{\uline}{\ul}{}
\usepackage{caption}
\usepackage{hyperref}
\usepackage{lineno}
\usepackage{ifpdf}
\usepackage{cite}
\usepackage{amsfonts}
\usepackage{amssymb}
\usepackage{lipsum,multicol,mathtools}

\begin{document}

\title{Implementation of Resistive Type Superconducting Fault Current Limiters in Electrical Grids: Performance Analysis and Measuring of Optimal Locations}
\author{X.~Zhang, H.~S.~Ruiz, Z.~Zhong, and T.~A.~Coombs
\thanks{Manuscript Submitted June 19, 2015.}
\thanks{X.~Zhang, Z.~Zhong, and T.~A.~Coombs are with the Electronics, Power and Energy Conversion Group, Department of Engineering, Electrical Engineering Division, University of Cambridge, 9 JJ Thomson Avenue, Cambridge, CB3 0FA, U.K. (e-mail: \href{mailto:xz326@cam.ac.uk}{xz326@cam.ac.uk})}
\thanks{H.~S.~Ruiz is with the Department of Engineering, University of Leicester, Leicester LE1 7RH, U.K., and also with the Electrical Engineering Division, Department of Engineering, University of Cambridge, Cambridge CB3 0FA, U.K. (e-mail: \href{mailto:dr.harold.ruiz@leicester.ac.uk}{dr.harold.ruiz@leicester.ac.uk}).}
\thanks{This work was supported by the Engineering and Physical Sciences Research Council (EPSRC) project NMZF/064. X. Zhang acknowledges a grant from the China Scholarship Council (No.~201408060080).}
}

%\author{X.~Zhang}
%\email{xz326@cam.ac.uk}
%\author{H. S. Ruiz}
%\email{hsr24@cam.ac.uk}
%\author{Z. Zhong}
%\email{zz272@cam.ac.uk}
%\author{T. A. Coombs}
%\email{tac1000@cam.ac.uk}

%\address{Electronics, Power and Energy Conversion Group, Department of Engineering, Electrical Engineering Division, University of Cambridge, 9 JJ Thomson Avenue, Cambridge, CB3 0FA, U.K.}
%

% The paper headers
%\markboth{IEEE Transactions on Power Systems,~Vol.~XX, No.~XX, Submitted June ~2015}%
%{X.~Zhang \MakeLowercase{\textit{et al.}}: Implementation of Resistive Type Superconducting Fault Current Limiters}

%
\maketitle

\begin{abstract}

In the past few years there has been a significant rise in the short-circuit current levels in transmission and distribution networks, it due to the increasing demands on power and the addition of sources of distributed generations. It leads to the need of integration of novel protection systems such as the superconducting fault current limiters (SFCLs), as the installation of these devices into the electric network aims to improve the overall system stability during normal and fault conditions, whilst the upgrading costs associated to the increasing demand for integration of renewables to the power grid are minimized. SFCL models on the electric distribution networks largely rely on the insertion of a step or exponential resistance that is determined by a predefined quenching time. However, beyond the framework of these models, the study of the performance, reliability, and location strategy for the installation of sole or multiple SFCLs in power grids still lacks of proper development leading to the utter need of comprehensive and systematic studies on this issue. In this paper, we expand the scope of the aforementioned models by considering the actual behaviour of a SFCL in terms of the temperature dynamic power-law dependence between the electrical field and the current density. Our results are compared with step-resistance models for the sake of discussion and clarity of the conclusions. Both SFCL models were integrated into a power system model built based on the UK power standard, and the impact of these protection strategies on the performance of the overall electricity network was studied. As a representative renewable energy source, a 90 MVA wind farm was considered for the simulations.  Three fault conditions have been simulated, and the figures for the fault current reduction predicted by both fault current limiting models have been compared in terms of multiple current measuring points and allocation strategies. Consequently, we have shown that the incorporation of the $E-J$ characteristics and thermal properties of the superconductor at the simulation level of electric power systems, is crucial for reliability estimations and optimal location of resistive type SFCLs in distributed power networks. Our results may help to the decision making by the distribution network operators about investment and promotion of the SFCL technologies, as a maximum number of SFCLs for different fault conditions and multiple locations has been determined. 

\end{abstract}

%\PACS{84.71.-b, 85.25.Am, 88.05.Bc, 88.05.Ec, 88.05.Lg, 88.80.Cd, 88.80.H, 88.50.Mp}
% 84.71.-b Superconducting high-power technology
% 85.25.Am	Superconducting device characterization, design, and modeling
% 88.05.Bc	Energy efficiency; definitions and standards
% 88.05.Ec	Renewable energy targets
% 88.05.Lg	Economic issues; sustainability; cost trends
% 88.80.Cd	Grid-connected distributed energy resources
% 88.80.H-	Electric power transmission
% 88.50.Mp	Electricity generation, grid integration from wind

\IEEEpeerreviewmaketitle

%%%%%%%%%%%%%%%%%%%%%%%%%%%%%%%%%%%%%%%
%%%%%%%%%%%% SECTION 1 %%%%%%%%%%%%%%%%
%%%%%%%%%%%%%%%%%%%%%%%%%%%%%%%%%%%%%%% 

\section{Introduction}\label{Section_1}

\IEEEPARstart{W}{ith} the tremendous increase on the electricity demands, the scale of both power grids and renewable energy generation systems is being expanded~\cite{Sung 2009}. Due to the persistent increase of conventional system generation and distributed generations (DGs), such as, photovoltaic plants, concentrating solar power plants, and wind farms, the likelihood of fault events capable to cause great and irreparable damage to a large set of electrical devices, or even system blackout, has been rapidly rising~\cite{Hartikainen2007}. This issue is now of major concern for the transmission system operators (TSOs), as increased fault current levels represents negative effects in terms of the reliability and security of the entire power systems~\cite{Alaraifi2009,Zhu2015}.

For the safe operation of power systems, various strategies for mitigating the fault current levels have been implemented in the power industry, such as, construction of new substations, split of existing substation busses, upgrade of multiple circuit breakers, and installation of high impedance transformers. Nevertheless, all these operational practices involve a not negligible degradation of the systems stability and their performance, what ultimately means the occurrence of significant economic losses and further investment~\cite{Kovalsky2005}. It is worth noticing that current limiting apparatuses, such as series reactors and solid state fault current limiters, are also widely used to reduce the fault levels in existing power grids. However, these devices insert impedance into the networks, permanently, and therefore cause a continuous voltage drop and power losses during normal operation~\cite{Ye2004}. 

Superconducting fault current limiters (SFCLs), are considered as the most promising alternatives to the conventional protective methods due to the remarkable features of the superconducting materials~\cite{Lee2013}. Specifically, during normal operation, SFCLs cause negligible voltage drop and negligible energy losses. However, when any of the critical limits defining the transition between the superconducting state to the normal state, such as critical temperature ($T_{c}$), critical current density ($J_{c}$), or critical magnetic field ($H_{c}$), is exceeded, the basic operational principle of a SFCL can be understood as an almost instantaneous quenching of the superconducting material from a negligible electric resistance to a highly resistive state. It gives to the SFCL the remarkable ability to limit faults even prior of attaining the first peak of a short-circuit current~\cite{Llambes2009}. In addition, a SFCL is capable to automatically restore to its superconducting state after the clearance of a fault, and its application requires no change in the existing network topologies.
 
According to the report provided by Morgan Stanley on Smart Grids, the potential market for fault current limiting devices may reach 5 billion dollars per year by 2030~\cite{MorganStanley2009}, but more comprehensive studies about the real impacts of installing SFCLs on electricity systems are still needed. In our previous work~\cite{Ruiz2015}, a systematic review on the successful field test and different existing numerical models of SFCLs has stated the viability of this technology. For performance simulation of SFCLs installed in real power grids, two simplified SFCL models were commonly used. The first approach was to model the SFCL as a step-resistance with pre-defined triggering current, quench time, and recovery time as in Ref.~\cite{Khan2011} and Ref.~\cite{Hwang2013}. This approach allows to consider a simplified scenario where no energy loss occurs during the superconducting state, and a high impedance is considered for modeling  its normal state, by ensuring that the SFCL responds to faults in an instantaneous fashion. However, it may lead to significant inaccuracies since the quenching and recovery characteristics depend on the thermal and electrical properties of the superconductors, which are both neglected in this case. On the other hand, the modeling of a SFCL into a power grid can also be simplified by using an exponential function for the dynamic resistance of the SFCL device,  in which the quenching action of the superconducting material is solely determined by time. This method has been previously implemented in Refs.~\cite{Park2010} and \cite{Park2011} in order to study the optimal location and associated resistive value of SFCLs for an schematic power grid with an interconnected wind-turbine generation system, finding that the installation of SFCLs can not only reduce the short-circuit current level but also, it can dramatically enhance the reliability of the wind farm. Compared with the Heaviside step function derived from the previous approach, this exponential resistance curve fits better with the real performance of the SFCL and furthermore provides aggregated computational benefits in terms of numerical convergence. Nevertheless, the SFCL characteristics including triggering current, quenching, and recovery time also have to be set before initializing the simulation, i.e., under this scenario also the physical properties of the superconductors are ignored. 

On the other hand, a more advanced model for resistive-type SFCL was presented in Ref.~\cite{Langston2005} where both the physical properties and the real dimensions of superconductors were considered. A similar model was then built by D. Colangelo et al.,\cite{Colangelo2013} in order to simulate the behavior of the SFCL designed in the ECCOFLOW project. Using this model the quenching action of the SFCL is no longer pre-defined. However, within these models the computational complexity is significantly increased, especially during large scale power network simulation. Hence, during performance simulation of SFCLs installed in power systems, it is rather important to study the necessity of considering thermal and electrical properties of superconducting materials, in order to be able to choose a better side between the tradeoff of computational complexity and model accuracy. Furthermore, it is worth noticing that for any of the adopted strategies, the research must ultimately addresses the finding of the optimal location of multiple SFCLs inside the power network, which as far of our knowledge it has been done by considering a maximum of just two SFCLs, which means that the cooperation and prospective need of more SFCLs remains as an open issue. 

In this paper we present a comprehensive study about the performance and optimal location analysis of resistive type SFCLs in realistic power systems, starting from the simplest consideration of a single step-resistance for the activation of the SFCL, up to considering the actual electro-thermal behavior of the superconducting component. We have simulated the performance of the SFCL under the scheme provided by two different models: (i) as a nonlinear resistance depending on time and, (ii) as a dynamic temperature-dependent model with the actual $E-J$ characteristics of the superconducting material. The applied power grid model which has interconnected dispersed energy resource was built based on the UK network standard. Through the simulations of the system behaviors under three fault conditions (two distribution network faults in different branches, and one transmission system fault), the optimum SFCL installation schemes were found from all the feasible combinations of SFCLs. In addition, a detailed comparison between the figures obtained for each one of the abovementioned cases has been performed, proving that the nonlinear resistor model is insufficient for an accurate estimation of the reliability figures and optimal location of the SFCL, as the complex thermal and electrical behaviors of the superconducting material during its transition to the normal state cannot be simplified to a single step-resistance.

This paper is organized as follows. Section~\ref{Section_2} introduces topologies and configurations of the power system. In Section~\ref{Section_3}, the two proposed resistive type SFCL models are described, and the comparisons between their performances are demonstrated. Then, the optimum installation schemes of the SFCL models are presented in Section~\ref{Section_4}, proving the importance of considering the thermal and electrical behaviors of superconductors. Furthermore, the need and effectiveness of implementing a switch strategy for improving the recovery characteristics of the SFCL is shown. Then, a comprehensive study about how the optimal location of multiple SFCLs can be determined in a large scale electrical grid is the purpose of Section~\ref{Section_5}.  Finally, the main conclusions of this paper are presented in Section~\ref{Section_6}.

%%%%%%%%%%%%%%%%%%%%%%%%%%%%%%%%%
%%%%%%%%               SECTION 2            %%%%%%%%%%%%%
%%%%%%%%%%%%%%%%%%%%%%%%%%%%%%%%% 

\section{Power System Model}~\label{Section_2}

%

%%%%%%%%%%%%%%%%
%%%%     FIGURE 1      %%%%%
%%%%%%%%%%%%%%%%
\begin{figure*}
\begin{center}
{\includegraphics[width=1.0\textwidth]{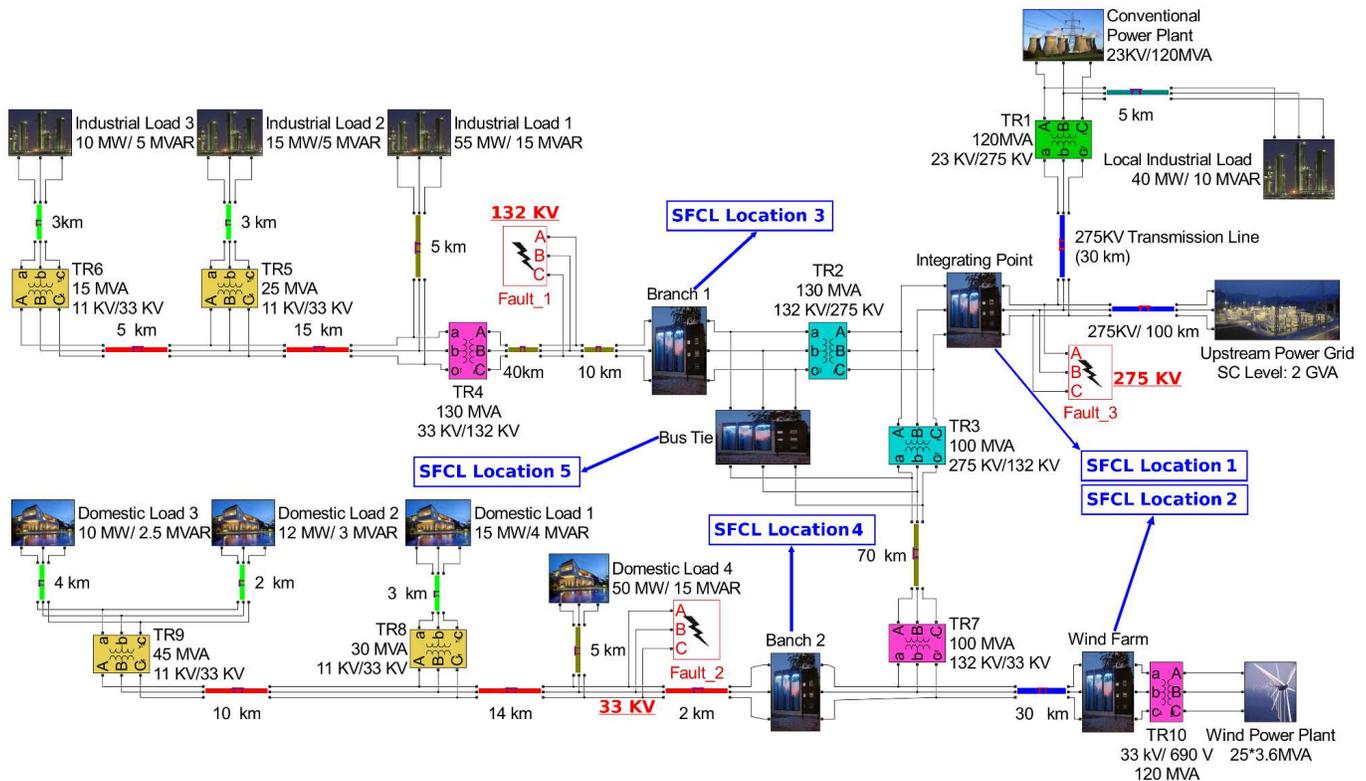}}
\caption{Power system model based on the UK grid standard as it is described in Section~\ref{Section_2}. Three prospective fault positions and five prospective SFCL locations are illustrated.}
\label{Figure_1}
\end{center}
\end{figure*}
%%%%%%%%%%%%%%%%

The modelled power system was built based on the UK network standard~\cite{UK-Networks}, and the interconnected wind power plant was designed according to the Rhyl Flats offshore wind farm located in North Wales, which has 25 wind turbines with a maximum rated output of 90 MVA \cite{Feng2010}. Fig.~\ref{Figure_1} shows the power system model developed in Matlab/Simulink/SimPowerSystems.

The power system has a 120 MVA conventional power plant emulated by a three-phase synchronous machine, which is additionally connected to a local industrial load of 40 MW located 5 km away from the main power plant. Afterwards, the voltage level is boosted from 23 KV to 275 KV by a step-up transformer (TR1), from where the conventional power plant is connected to an upstream power grid rated with a short circuit level of 2GW through 130 km long distributed-parameters transmission line. Then, the 275 KV high-voltage transmission system is split into two distribution networks. First, after voltage level stepped down to 33 KV by substations TR2 and TR4, the upper branch (industrial branch) supplies power to three industrial loads which rated power are 55 MW, 15 MW, and 10 MW, separately. Likewise, the lower branch (domestic branch) is also connected to two step-down substations TR3 and TR7, with 70 km distance between them. The role of these two substations is reduce the voltage of the lower sub-grid to 33 KV, as it is the same voltage level rated by the 90 MVA wind farm after being boosted by TR10. This offshore wind power plant is composed of twenty-five fixed-speed induction-type wind turbines each having a rating of 3.6 MVA, and it is located 30 km away from its connecting point with the lower distribution network. After integration, the lower branch and the wind farm together provide electric energy to four domestic loads with rated power of 50 MW, 15 MW, 12 MW and 10 MW, separately. In addition, it is worth mentioning that the industrial branch and the domestic branch are connected through a bus-bars coupler, and the power system is balanced in a way that the current flowing through the bus-tie is only of a few amperes during normal operation.

It is generally accepted that the three-phase (symmetric) short-circuit fault provokes the highest fault current among all possible faults, since it will cause the most drastic decrease of the system impedance. In order to ensure safe operation, the maximum current and electrodynamic withstand capabilities of electrical equipment are primarily designed according to this situation. Therefore, it is essential to simulate the behavior of the power system under three-phase short-circuit fault. The symmetric faults were initialized at three potential locations marked as Fault 1 (132 KV), Fault 2 (33 KV) and Fault 3 (275 KV), which represent prospective faults occurring at the industrial branch, the domestic branch, and the transmission system, respectively (see Fig.~\ref{Figure_1}). Five positions for the installation of SFCLs are proposed as shown in Fig.~\ref{Figure_1}, namely at: (i) the integrating point between the conventional power plant and the upstream power grid (Location 1), (ii) the interconnection between the wind farm and the port of domestic branch (Location 2), (iii) the industrial loads branch (Location 3), (iv) the domestic loads branch (Location 4), and (v) the bus-tie coupling the two distribution networks (Location 5). 

It is worth mentioning that due to insufficient margin of short-circuit capacity in some MV grids, nowadays the dispersed power plants have to be directly connected to the HV grids through expensive generator transformers. However, this considerable investment could be avoided by installing the SFCL at the port of the distributed generation (Location 2)~\cite{Neumann2006}. Furthermore, the SFCL as bus-bar coupler (Location 5) is also one of the most promising locations for the installment of a SFCL because it would lead to lower losses whilst concomitantly enable parallel operation of transformers with doubled short-circuit capacity. It ultimately results in lower voltage drops and an overall improved stability of the power system. Moreover, the installation of a SFCL at Location 5 could bring about substantial economic benefits as it may allow the direct connection of harmonic polluting loads (e.g. arc furnace) and high loadings to the MV bus-bars, which otherwise have to be connected to the HV grid through an appropriate transformer that normally requires of considerable investment~\cite{Neumann}.

%%%%%%%%%%%%%%%%%%%%%%%%%%%%%%%%%
%%%%%%%%               SECTION 3            %%%%%%%%%%%%%
%%%%%%%%%%%%%%%%%%%%%%%%%%%%%%%%% 

\section{Resistive Type SFCL Models}~\label{Section_3}

Several designs of SFCLs have been developed in the past, which can be categorized into resistive type, inductive type, and hybrid type devices. In this paper, we focus on the resistive type due to its compactness and stability compared with other designs~\cite{Eckroad2009}. Two three-phase SFCL models, namely non-linear resistance model and $E-J$ Power Law based model were built, and their operating principles are presented in this section. Both models are constructed by three identical single-phase SFCL modules, since each phase of the SFCL would only be triggered by the current flowing through its own phase. Specifically, unbalanced faults could only quench one or two phases of the entire SFCL. Furthermore, under condition of symmetric faults of the grid, each phase of the SFCL will quench slightly asynchronously within the first cycle of the fault current, leading to an instantaneous unbalance between the phases. Hence, for all types of faults and at diverse locations, independent modules for each one of the three phases have to be considered in order to allow an accurate simulation of the SFCL's effects on the overall power grid~\cite{Blair2012}. Below, the aforementioned models are described in further detail.

%%%%%%%%         Subsection 3.1            %%%%%%%%%%%%

\subsection{Step resistance SFCL model}~\label{SubSection_3_1}

Current limiting performance of the developed step resistance model is dominated by five predefined parameters of the SFCL: (i) triggering current, (ii) quenching resistance, (iii) quenching time, which has been assumed equals to 1~ms accordingly with Refs.~\cite{Sung 2009} and \cite{Alaraifi2009}, (iv) a normal operating resistance of $0.01~\Omega$ and, (v) a recovery time of 1~s. The values of triggering current and quenching resistance are not provided in this section since they vary with the locations of SFCL.

%%%%%%%%%%%%%%%%
%%%%     FIGURE 2      %%%%%
%%%%%%%%%%%%%%%%
\begin{figure*}[t]
\begin{center}  
{\includegraphics[width=1.0\textwidth]{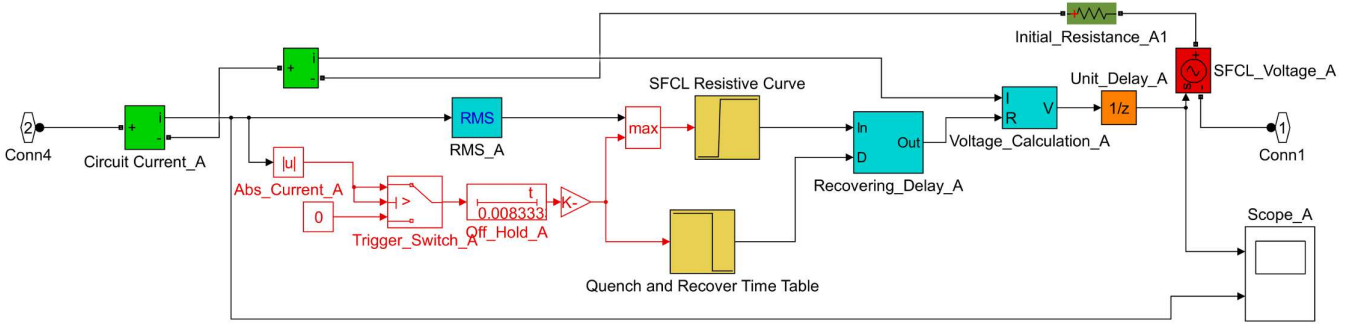}}
\caption{One Phase of the step resistance SFCL model.}
\label{Figure_2}
\end{center}
\end{figure*}
%%%%%%%%%%%%%%%%

The structure of the step resistance model is illustrated in Fig.~\ref{Figure_2}. The operating principle of this model can be summarized as follows: first, the SFCL model calculates both the absolute and RMS values of flowing current. If both values are lower than the triggering current, the model will consider the SFCL in superconducting state and insert normal operating resistance (0.01 $\Omega$) into the grid. On the contrary, if either the absolute value or the RMS value of a passing current exceeds the triggering current level, the output resistance will be increased to the quenching resistance after the predefined quenching time. Lastly, if current flowing through the SFCL model falls below the triggering current due to the clearance of the fault, the SFCL will restore its superconducting state after the recovery time. 

On one hand, compared with the instantaneous current, the RMS calculation always has certain delay due to the nature of integral operation, which may cause the SFCL incapable to limit the first peak of fault current. On the other hand, if the quenching activation solely depends on the absolute value of instantaneous current, it could lead to fault switching when the flowing AC current close to zero in one period. Cooperation of these two different strategies could enable the SFCL model to quench and recover not only in a fast but also in a stable fashion.

%%%%%%%%%%%% Subsection 3 2 %%%%%%%%%%%%%%%%
\subsection{SFCL model with E-J power law characteristics and dynamic temperature}~\label{SubSection_3_2}

The fundamental operation principle of the resistive type SFCLs relies in the almost immediate insertion of a high resistance into the power grid, once the superconducting material is fully quenched. Therefore, the sudden change in the SFCL resistance is mainly due to the electrical properties of the superconductor, which can be macroscopically simplified into the so-called $E-J$ power law~\cite{Rhyner1993}. One commonly used method of the Bi2212 based SFCL modeling is to subdivide the $E-J$ characteristic of the superconductors into three sub-regions: superconducting state, flux flow state, and normal conducting state~\cite{Langston2005,Blair2012,Paul2000,Aly2012}. All three sub-regions follow different power laws, combination of which forms the entire $E-J$ characteristics of the SFCL as follows:

\begin{IEEEeqnarray}{rCl}
E(T,t)= 
\begin{cases}
    E_{c}\left(\dfrac{J(t)}{J_{c}(T(t))}\right)^{n} ,  \\ \hspace{0.5cm} \text{for $E(T,t)<E_{0}$ and $T(t)<T_{c}$}.\\
    E_{0}\left(\dfrac{E_{c}}{E_{0}}\right)^{m/n}\left(\dfrac{J_{c}(77 K)}{J_{c}(T(t))}\right)\left(\dfrac{J(t)}{J_{C}(77 K)}\right)^{m} , 
    \\ \hspace{0.5cm} \text{for $E(T,t)>E_{0}$ and $T(t)<T_{c}$}.\\
    \rho(T_{C})\dfrac{T(t)}{T_{C}}J(t)  ,
     \\  \hspace{0.5cm} \text{for $T(t)>T_{c}$}.
\end{cases}
\label{Eq_1}
\end{IEEEeqnarray}
where,
\begin{equation}
J(T,t)=J_{c}(77K)\dfrac{T_{c}-T(t)}{T_{c}-77},~~~~  for~~J\textgreater J_{c}~~and~~T(t)\textless T_{c}  ~ .
\label{Eq_2}
\end{equation}

For the modeling of the superconducting state we use the power index $n=9$ in accordance with Refs.~\cite{Buhl1997,Paul1993,Herrmann1996,Bock2005,Noe2001}, and $m=3$ for the flux flow state as it has been observed that this change in the power law from $n$ to $m$ shows good agreement with the experimental data reported in Refs.~\cite{Paul2001} and \cite{Chen2002},  automatically including the effect of the self-induced magnetic field. 

On the other hand, we apply $\rho(T_{c}) = 7\times 10^{-6} \Omega$ and consider the normal conducting state resistivity as a linear function of temperature when $T(t)> T_{c}$.\cite{Elschner2001} This approximation is considered to be a reasonable assumption as proved in Ref.~\cite{Liu2001}. Furthermore, the relationship between the critical current density and the temperature is also set to be linear as in Eq.(\ref{Eq_2}), as it has been proved by S. Kozak et al. for the specific case of Bi2212 compunds~\cite{Kozak2005}. In addition, for completing the SFCL model, a CuNi alloy $(\rho=40\mu\Omega\cdot m)$ resistor is connected in parallel with the superconductor on the basis of the project disclosed in Ref.~\cite{Rettelbach2003}. This shunt resistance can protect the superconducting material from being damaged by hot spots that are developed under limiting conditions, and furthermore prevents over-voltages that possibly appear if the quench occurs too rapidly~\cite{Noe2007,Bock2004}.

Finally, the temperature developed by the superconducting material is calculated based on the intensity of flowing current, the heat capacity of the Bi2212 bulk, and its thermal resistance, in a constant bath of liquid nitrogen at $77 K$. Then, under assumption that the superconducting composite is homogeneous, the thermal modeling of the SFCL considers the first order approximation of the heat transfer between the superconductor and the liquid nitrogen bath as follows:
\begin{eqnarray}
R_{SC} & = & \dfrac{1}{2\kappa\pi d_{SC}l_{SC}}  ~ , \label{Eq_3} \\
C_{SC} & = & \dfrac{\pi d_{SC}^{2}}{4}l_{SC}c_{v}  ~ , \label{Eq_4} \\
Q_{generation}(t) & = &I(t)^{2}\times R_{SFCL}(t) ~ , \label{Eq_5} \\
Q_{cooling}(t) & = & \dfrac{T(t)-77}{R_{SC}} ~ , \label{Eq_6} \\
\end{eqnarray}
where $R_{SC}$ stands for the thermal resistance from the superconducting material to its surrounding coolant, $C_{SC}$ is the superconductor heat capacity that in our case corresponds to the specific heat of Bi2212 ~\cite{Meerovich2007}, $c_{v}=0.7\times 10^{-6} J/(m^{3}\cdot K)$, and
\begin{eqnarray}
T(t)=77+\dfrac{1}{C_{SC}}\int_{0}^{t}[Q_{generated}(t)-Q_{generated}(t)]dt ~ . \label{Eq_7}
\end{eqnarray}

For the SFCL installed at different locations, the superconductor is generically modeled as a cylindrical wire whose length $l_{SC}$ is adjusted in order to limit the prospective fault current to the desired level, and the diameter $d_{SC}$ is regulated to ensure that the SFCL not only remains into the superconducting state during normal operation, but also quenches within a few milliseconds once a short-circuit fault occurs at some location into the grid. In practice, despite the wire diameter cannot be modified after fabrication, one can connect several wires in parallel to achieve the expected current limiting performance~\cite{Blair2011}.

%%%%%%%%%%%%%%%%%%%%%%%%%%%%%%%%%
%%%%%%%%               SECTION 4            %%%%%%%%%%%%%
%%%%%%%%%%%%%%%%%%%%%%%%%%%%%%%%% 

\section{Performance and impact of SFCLs on the grid network stability}~\label{Section_4}

In order to compare the fault current limitation properties of the two SFCL models, a three-phase to ground fault with negligible fault resistance was initialized at the domestic loads network (Fault 2) in the grid model shown in Fig.~\ref{Figure_1}, when a single SFCL is installed next to the fault position (Location 4). 

\begin{figure*}[t]
\begin{center}
{\includegraphics[width=1.0\textwidth]{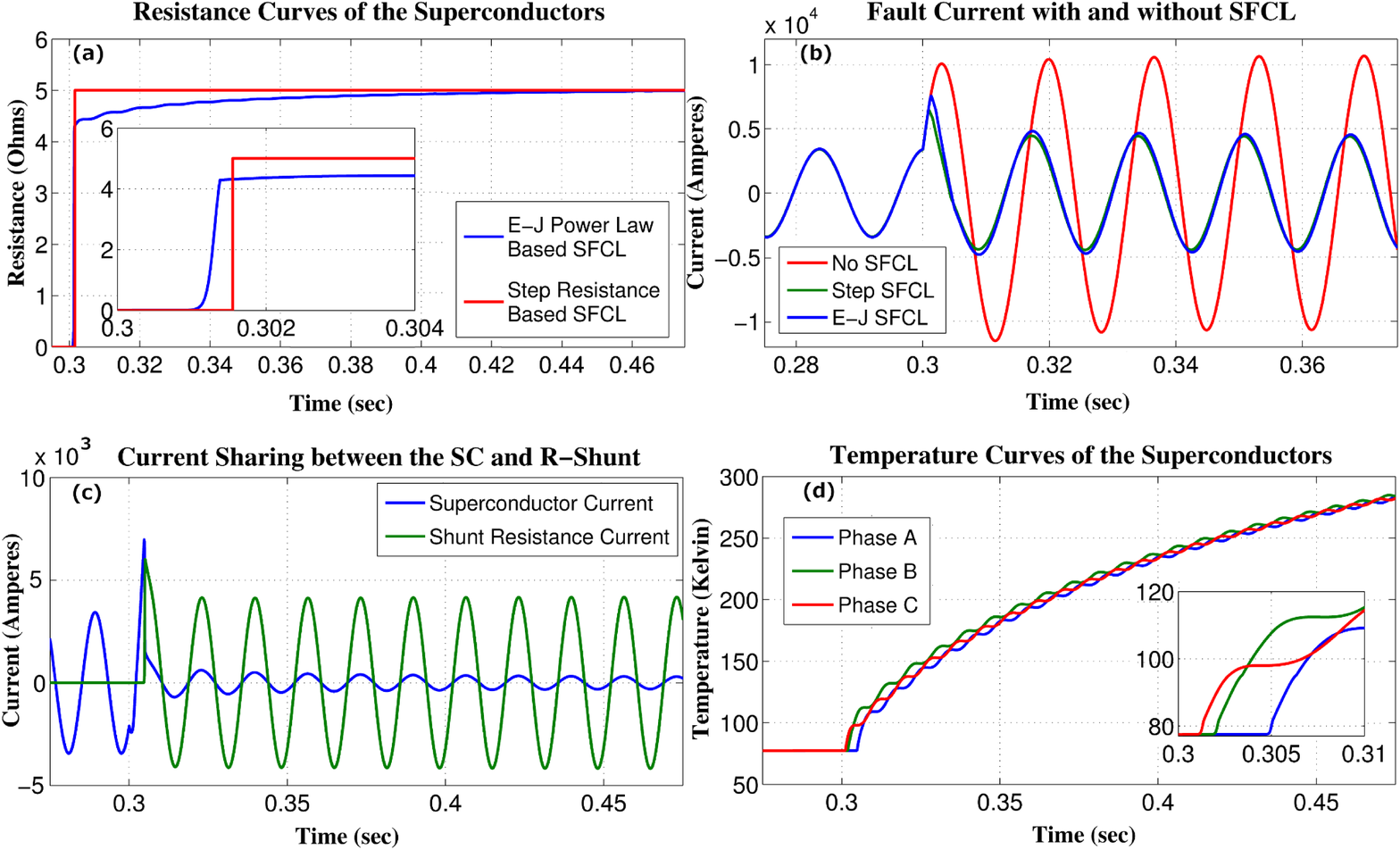}}
\caption{(Color Online) Performance comparison between the step SFCL model and the $E-J$ power law based SFCL model. The displayed insets in subplots (a) and (d) are measured in the corresponding units of the main plot. For a detailed discussion of the results presented, please refer to Section~\ref{Section_4}.}
\label{Figure_3}
\end{center}
\end{figure*}

Fig.~\ref{Figure_3}(a) illustrates that the step resistance model and the $E-J$ power law based model both respond almost simultaneously to the occurrence of a short-circuit fault. However, since in reality the SFCL needs $\sim2$ ms to fully quench due to its $E-J$ characteristic and dynamic temperature (Fig.~\ref{Figure_3}(d)), the first peak reduction gained onto the step resistance model is overestimated by 11\% (7.6~kA and 6.5~kA for the two SFCL models, respectively. 10~kA without SFCL), as shown in Fig.~\ref{Figure_3}(b). In addition, the shunt resistor diverts the major portion of the fault current after the superconductor develops its normal state (Fig.~\ref{Figure_3}(c)). Therefore, the shunt resistance effectively lowers the thermal stress on the HTS wire, simultaneously preventing damages by overheating, whilst the recovery time is reduced~\cite{Morandi2013}.

%%%%%%%%%%%%%%%%
%%%%     FIGURE 3      %%%%%
%%%%%%%%%%%%%%%%
%%%%%%%%%%%%%%%%

Initial tests without integration of the SFCL model have confirmed that the power system operates at rated state during normal operation. Then, under occurrence of three-phase to ground faults at Fault-1, Fault-2 and Fault-3 (see Fig.~\ref{Figure_1}), the short-circuit currents were measured at the integrating point (Location 1), wind farm (Location 2), branch 1 (Location 3) and branch 2 (Location 4), as shown in Fig.~\ref{Figure_1}.  

The instantaneous fault current can be described by the following equation~\cite{Saadat1999}:
\begin{eqnarray}
i_{k}= & \underbrace{I_{pm}sin(\omega t+\alpha-\beta_{kl})}_{\text{periodic component}}+\nonumber\\ & \underbrace{[I_{m}sin(\alpha-\beta)]-I_{pm}sin(\alpha-\beta_{kl})]e^{-\dfrac{t}{\tau_{k}}}}_{\text{aperiodic component}} ~\, ,
\label{Eq_8}
\end{eqnarray}
where $I_{m}$ is the amplitude of the rated current of the power grid, $\phi$ and $\phi_{kl}$ represent the impedance angles before and after the fault, respectively; $\alpha$ defines the fault inception angle, $I_{pm}$ states the magnitude of periodic component of the short-circuit current which is defined by the source voltage and circuit impedance, and finally $\tau_{k}$ stands for the time constant of the circuit. Thus, the fault currents achieve their maximum values when $\alpha-\beta_{kl}=\pi (n+1) / 2$ with $n\in \mathbb{Z}$. Therefore, in order to consider the most hazardous fault scenarios that could occur into the grid, the considered short-circuit faults are initialized under this condition, when one or multiple SFCLs are installed to the network.
\subsection{SFCL's impact on generation and voltage stability}~\label{SubSection_4_1}
\begin{figure*}[t]
\begin{center}
%\hspace*{-2.5cm}   
{\includegraphics[width=1.0\textwidth]{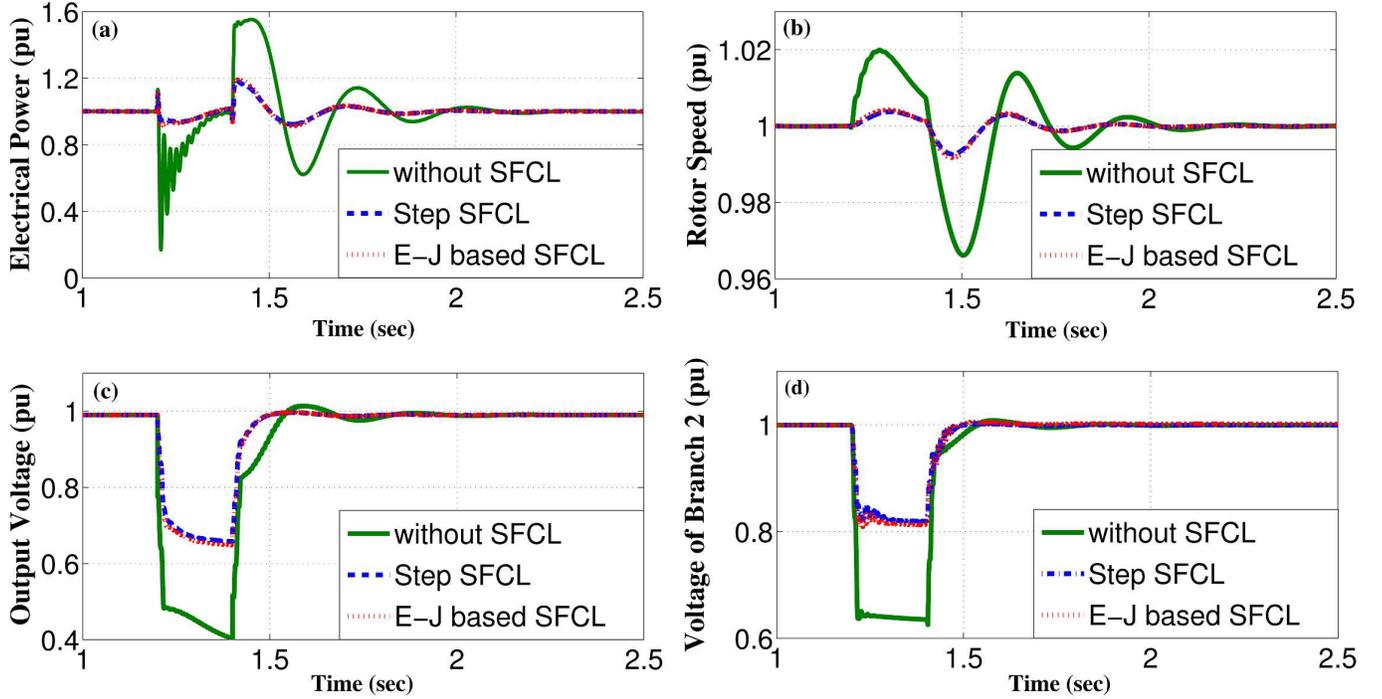}}
\caption{(Color Online) Generator parameters and voltage of branch 2 in response to a 200 ms three-phase to ground fault at branch 1.}
\label{Figure_4}
\end{center}
\end{figure*}

When a 200 ms three-phase to ground fault is applied at the industrial branch, also called branch 1 in Fig.~\ref{Figure_1} (Fault-1), after a time period of 1.2 s within normal operating conditions, the response of the output electrical power, rotor speed, and terminal voltage for the conventional power plant (23~KV~/~120MVA), and the voltage output at the domestic branch (Branch 2) are shown in Fig.~\ref{Figure_4}. Initially we have to consider the power system operation without the insertion of SFCLs. Under this scenario, the output electrical power drops sharply to 0.15 $pu$ just after the fault incident (Fig.~\ref{Figure_4}(a)),  whilst on the other hand, governors of the power plant such as steam and hydro still contribute with the same mechanical power to the rotors. Therefore, a  rapid acceleration of the rotors occurs due to this power imbalance, as shown in Fig.~\ref{Figure_4}(b). Furthermore, the generators have to oscillate violently for another second before it could be stabilized again after the short-circuit fault being cleared at 1.5 s.

%%%%%%%%%%%%%%%%
%%%%     FIGURE 4      %%%%%
%%%%%%%%%%%%%%%%
%%%%%%%%%%%%%%%%

However, when a SFCL is installed at Branch 1 (Location 1), its high resistance state facilitates the SFCL to dissipate the excess generator power during the fault condition, hence improving the energy balance of the system and reducing the variation of the rotor speed effectively. Furthermore, with consideration of the conventional equal-area criterion for stability issues~\cite{Sung 2009}~\cite{Kundur1994}, the SFCL could improve the damping characteristics of generator speed, system frequency, as well as the system current, since the insertion of the high resistance into the grid would significantly increase the damping ratio. Moreover, due to the short-circuit fault of Branch 1, a sharp voltage drop can be seen at both the power plant terminal (0.5~$pu$) and the non-faulted Branch 2 (0.35~$pu$), as it can be seen in Figs.~\ref{Figure_4}~(c) \& (d). Then, by introducing a SFCL which acts as a voltage booster, the observed voltage dips are mitigated by 40\% and 50\%, respectively. This improvement allows the healthy parts of the system (without the fault inception) to be less affected and, makes the use of a  SFCL a reliable fault ride-through scheme.

\subsection{Current limiting performance versus maximum normal resistance}~\label{SubSection_4_2}

\begin{figure*}
\begin{center}
{\includegraphics[width=1.0\textwidth]{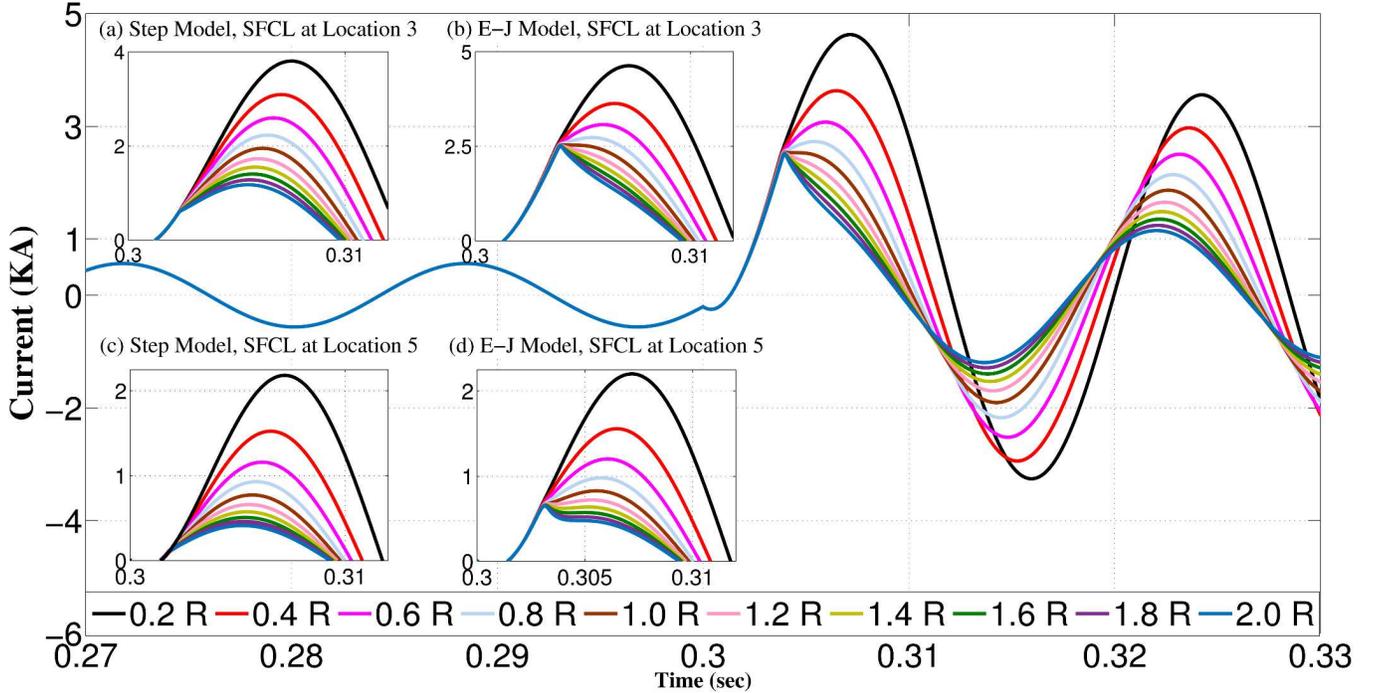}}
\caption{(Color Online) Current curves of phase A under branch 1 fault (Fault 1) when SFCL resistance increases from 0.2 R to 2.0 R.}
\label{Figure_5}
\end{center}
\end{figure*}

A 200 ms short-circuit fault was initiated in Branch 1 (Fault 1) in order to study the relationship between the current limiting performance of SFCL and the maximum normal resistance. Firstly, without the protection of the SFCL, simulation results have shown that the first peak of current flowing into Branch 1 reaches $\sim3.8$~kA, which means $\sim 6.8$ times higher than the rated value (560 A). Then, after installing the SFCL into the power grid, a considerable reduction of the fault current is observed as shown in Fig.~\ref{Figure_5}. The insets (a) and (b) on this figure illustrate the variation of the limited current when the two SFCL models (step model, and $E-J$ power law based model) are integrated at Branch 1 (Location 3). 

With the normal resistance of the step SFCL model increasing from 0.2~R to 2~R ($R=30\Omega$), the peak value of the fault current gradually decreases from $\sim3.8$~kA to $\sim1.2$~kA with a small displacement of the peak values. However, in the case of the $E-J$ power law based SFCL model, a noticeable \textit{kink} appears at about 2.5~kA, when the maximum resistance of the SFCL is greater than $1~R$. This \textit{kink}, distinctive of the $E-J$ power law based model studies, can be interpreted as the threshold value for the maximum reduction of the fault current, i.e., for instance, in the present case of study the greatest current reduction that the $E-J$ model could achieve is $\sim1.3$~A lower than the result acquired from the step model. This significant difference between the performances of the two SFCL models is therefore, caused by the superconductor's electrical characteristics and thermal properties. 

After the current flowing through the SFCL exceeds the critical limit of the superconducting material, the developed resistance rises exponentially with a factor $n$ depending on the material choice, $n=9$ in the case of Bi2212, and finally enter to the normal conducting state. This transition occurs within 1 ms and thus enables the SFCL to limit the first peak of the fault current. Nevertheless, certain amount of time is always demanded for heat accumulation, quenching process, and resistance rise. We have determined that on the instant that the \textit{kink} appears, the resistance developed by SFCLs of different sizes only present small differences, so the current curves approximately overlap at about 2.5~kA. Afterwards, the SFCLs that have higher capacities keep increasing their resistance, leading to sudden drops of currents, accordingly. Similar behavior is also observed when the SFCl is located ot other different positions, such as at bus-tie (Location 5), as shown in Fig.~\ref{Figure_5}~(c)~\&~(d).

The analysis above demonstrates that when the internal performance of the SFCL is emulated by using the $E-J$ power law based model, this strategy is more suitable for performing current limiting studies, because the step resistance model may overestimate the actual performance of the SFCL significantly. Furthermore, the use of the $E-J$ power law based model allows to determine in a more accurate way the optimal normal resistance of the SFCL devices given the occurrence of the \textit{kink} phenomenon. In terms of economic figures, it represents a very valuable result for the distribution operators as it allows to state a maximum threshold for the estimation of the actual need for enlarging the SFCL's capacity. Thence, excessive investments can be avoided, as beyond this threshold no further reduction of fault currents can be achieved.

%%%%%%%%%%%%%%%%
%%%%     FIGURE 5      %%%%%
%%%%%%%%%%%%%%%%
%%%%%%%%%%%%%%%%

\subsection{Bypass switch for improving the recovery characteristics of the SFCL}~\label{SubSection_4_3}

The passive transition of the superconducting material and the high normal resistance enables the SFCL to considerably limit the prospective fault current even before the first peak. However, when modeling the SFCL by considering its physical properties into the $E-J$ power law based model, in some cases the recovery characteristics need to be improved because the SFCL may need several minutes to restore to the superconducting state under load conditions. For instance, if a single SFCL located at the domestic branch fully quenches to limit a fault, it will then cost more than 300 seconds to recover from its resistive state after the fault cleared.

%%%%%%%%%%%%%%%%
%%%%     FIGURE 6      %%%%%
%%%%%%%%%%%%%%%%
\begin{figure*}[t]
\begin{center}
%\hspace*{-2.5cm}   
{\includegraphics[width=1.0\textwidth]{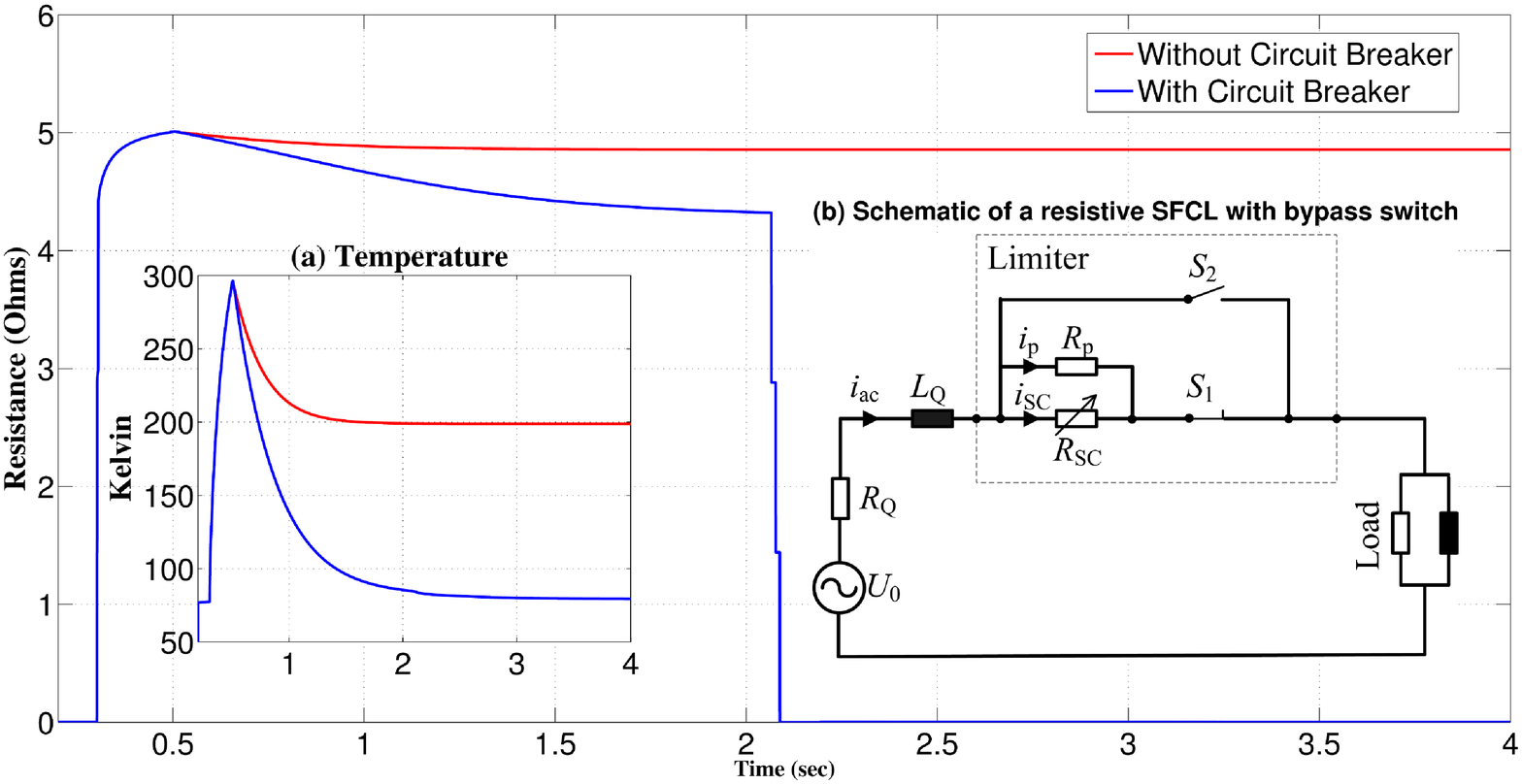}}
\caption{(Color Online) Schematic of a resistive SFCL with parallel bypass switch.}
\label{Figure_6}
\end{center}
\end{figure*}
%%%%%%%%%%%%%%%%

Therefore, in order to decrease the recovery time of the SFCL we have connected a bypass switch parallel to the superconductor and the shunt resistance as shown in Fig.~\ref{Figure_6} (b), which has been already proved to be a feasible and reliable strategy~\cite{Melhem2011}. Under the situations where the SFCL can recover fast enough under load conditions, as for instance when the SFCL is installed as bus-tie coupler (Location 5) then, the switch $S1$ remains closed after the fault is cleared. On the contrary, if the SFCL cannot be automatically recovered within a short time then, the switch $S2$ can be closed and the switch $S1$ is instantaneously opened to quickly disconnect the superconducting element from the system. Thereby, it allows the superconductor to start its recovery process without further accumulation of heat. Thus, by implementing this scheme, the recovery time of SFCLs could be substantially reduced to a few seconds in most of the cases, it depending on the dimensions of the superconducting elements and the temperature before actioning the bypass switches.

Fig.~\ref{Figure_6} shows the recovery characteristics of the SFCL at Location 2 after encountering a 0.2 s three-phase to ground fault at the domestic branch (Fault 2). Without applying the bypass switch, certain amount of current will continue passing through the SFCL after clearance of the fault. This flow of current keeps generating heat inside of the superconductor, what significantly slows down the decrease of temperature, and hence delay the recovery of the SFCL by more than five minutes. However, with a properly designed control scheme, the $E-J$ model can open the switch $S1$ and close the switch $S2$ at the moment that the fault ends, thus transferring the current to the $S2$ branch and isolating the superconducting material in order to help with the cooling process and reduce cryogenics investments. In fcat, by using this method we have determined that the recovery time can be reduced to less than 1.6 s without affecting the normal operation of the power grid. Then, after the superconductor is restored to its superconducting state, the switches $S1$ and $S2$  act again, in order to prepare the SFCL for the next fault.

%%%%%%%%%%%%%%%%%%%%%%%%%%%%%%%%%
%%%%%%%%               SECTION 5            %%%%%%%%%%%%%
%%%%%%%%%%%%%%%%%%%%%%%%%%%%%%%%% 

\section{Optimal allocation strategies for the installment of SFCLs}~\label{Section_5}

In order to obtain an accurate estimation of the optimal strategies for installation of SFCLs, all possible SFCL combinations according to the five proposed locations depicted in Fig.~\ref{Figure_1} were analyzed for three different fault points. Thus, in total we have to consider 31 allocation strategies which include five different schemes for the installment of a single SFCL at Locations 1 to 5, separately, 10 dual combinations of SFCLs, 10 further combinations of three SFCLs, as well as five scenarios where four SFCLs are working together, and finally the cooperation between all five SFCLs. The current signals at both wind farm terminal (Location 2) and the integrating point of conventional power plant and upstream power grid (Location 1) were measured for all three fault conditions (Fig.~\ref{Figure_1}). Moreover, we also analyzed the current injection of industrial branch (Location 3) and domestic branch (Location 4) when faults happen at the two networks, Fault 1 and Fault 2, respectively.  It has to be noticed that the present study has been done in a very exhaustive fashion, as it is the only way to assure that the optimal allocation strategy for multiple combinations of SFCL is actually adequate for the consideration of realistic power networks. It is also worth to mention that in the present paper we do not present the results for the measured current at the industrial branch when the Fault 2 or Fault 3 occurs, because based on the analysis of the system impedance change, the magnitude of the current flowing into the industrial branch is actually reduced by the two faults to levels lower than the normal current, i.e., at this point the SFCL does not need to be triggered to protect this branch. The same argument applies to the domestic branch under Fault 1 and Fault 3 conditions. Our results are presented below in terms of the single or multiple SFCL strategies.

\subsection{Single SFCL installation}~\label{SubSection_5_1}

%%%%%%%%%%%%%%%%
%%%%     FIGURE 7      %%%%%
%%%%%%%%%%%%%%%%
\begin{figure*}[t]
\begin{center}
%\hspace*{-2.5cm}   
{\includegraphics[width=1.0\textwidth]{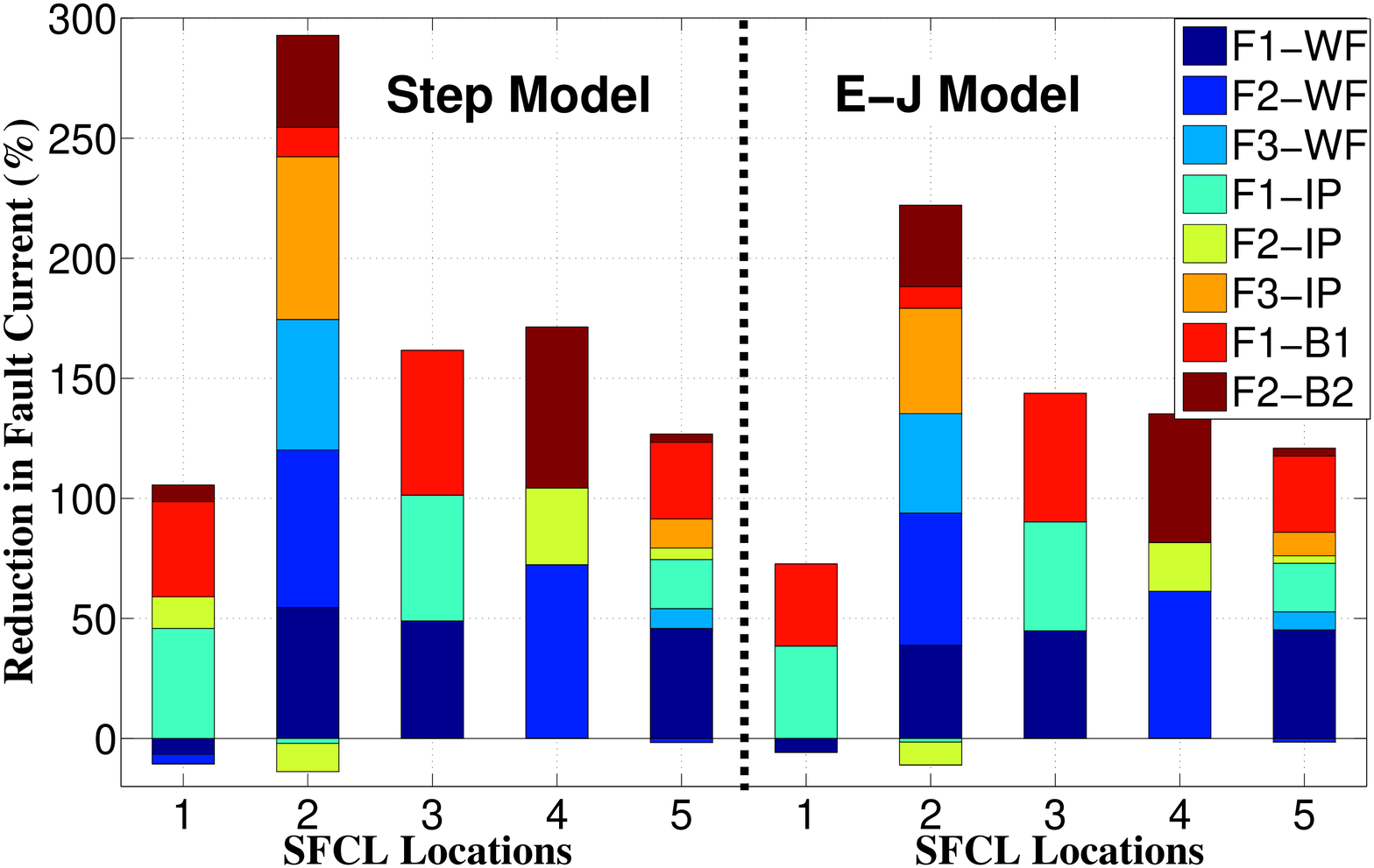}}
\caption{(Color Online) Reduction in first peaks of fault currents achieved by single SFCL at different locations.}
\label{Figure_7}
\end{center}
\end{figure*}
%%%%%%%%%%%%%%%%

Fig.\ref{Figure_7} shows the reduction in fault current under the three fault conditions illustrated in Fig.~\ref{Figure_1} when a single SFCL is installed at at the referred locations (Location 1 to  5). The size of the superconductor which has to be defined into the $E-J$ power law based model for the simulation of the SFCL, was systematically adjusted so that it allows to define the same maximum resistance with the step model for the sake of comparison. Thus, it is worth noting that when the step resistance model is considered, the maximum reduction of the fault current is overestimated in comparison with the more realist $E-J$ power law based model, i.e., for all the 5 SFCL locations the first peak of the fault current was always found to be lower in the first case. The reason to this difference is that, once the current exceeds the critical value of the superconducting material, the SFCL described by the step model directly jump to the maximum resistance after the pre-defined response time, whilst in the $E-J$ power law based model the dynamic increase of the resistance depends not only on the passing current, but also on the temperature of the superconductor. Therefore, under the $E-J$ power law model the SFCL cannot gain its maximum rated resistance before the first fault peak is reached, which leads to a relatively lower reduction of the fault current in about 20\%.

The simulations performed, based on both SFCL models, generally show a negative impact on the reduction of the fault peak at certain integration points when the SFCL is installed at Location 1 or Location 2, i.e., on these cases the fault current may actually increase by the insertion of a SFCL. In more detail, when the SFCL is installed besides the wind farm (Location 2), a sudden increase of fault current flowing through the integrating point under the  Fault 2 condition (at the domestic branch) is caused by the abrupt change of power system's impedance. This SFCL enters the normal state reducing the current output of the wind farm due to its rapid rise in resistance and hence, the conventional power plant and the upstream power grid are forced to supply higher current to the faulted branch. Similar behavior is obtained under the fault conditions F1 and F2 when the SFCL is installed at Location 2  and the current is measured at the integrating point (see Figs.~\ref{Figure_7}~\&~~\ref{Figure_1}). Furthermore, it should be noticed that when a single SFCL is installed at Location 1 (integrating point) following the $E-J$ power law based model, the SFCL can only limit the fault current in two cases whilst with the simplified step-resistance model the benefits of the SFCL can be overrated as it leads to a positive balance in up to four different fault conditions. This highlights the importance of finding a suitable optimal allocation strategy for the SFCLs under a wide number of fault conditions and, the need of considering adequate physical properties for the electro-thermal dynamics of the superconducting materials, which ultimately try to fill the gap between the acquired scientific knowledge and the demand of more reliable information from the standpoint of the power distribution companies. In fact, in this terms, we can conclude that when the measuring is taken as example at the integrating point shown in Fig.~\ref{Figure_1}, and a fault occurs at the domestic branch (Fault 2), the current produced by the conventional power plant and the upstream power grid flow through 70 km of transmission lines and two transformers (TR3 and TR7), whose resistance together is almost double compared with the resistance between the wind farm and the fault point (30 km transmission line and one transformer TR10). Therefore, most of the fault current flowing into the domestic branch is supplied by the wind farm, and the passing current at the integrating point only increases from 500 A to 830 A. Thus, when a SFCL is installed at the integrating point (Location 1), and the step resistance model is considered, the SFCL can be seen as an effective mechanism by reducing the current to about 700~A with a proper grading. However, in the reality, this reduction of the fault current cannot be achieved because of the transient states of the superconducting material during normal and fault conditions (see Eq.~\ref{Eq_1}). Furthermore, generally certain gap between the triggering current and the rated current of power system needs to be guaranteed, i.e., if the diameter of the superconducting material is adjusted such that it enables the SFCL to limit the first peak of the fault current from 830~A to 700~A, under the framework of the step resistance model the SFCL's resistance during normal operation would have to increase from 0.01 $\Omega$ to 0.14 $\Omega$. This significant rise in resistance is unacceptable considering that the thermal loss can then reach $\sim35$~KW, causing severe burden on the cooling system. Therefore, despite the very convenient simplicity of the step resistance model, determining the actual scope from derived studies based upon this model may result in too ambiguous approximations for the distribution operators.

%%%%%%%%%%%%%%%%
%%%%     FIGURE 8      %%%%%
%%%%%%%%%%%%%%%%
\begin{figure*}[t]
\begin{center}
%\hspace*{-2.5cm}   
{\includegraphics[width=1.0\textwidth]{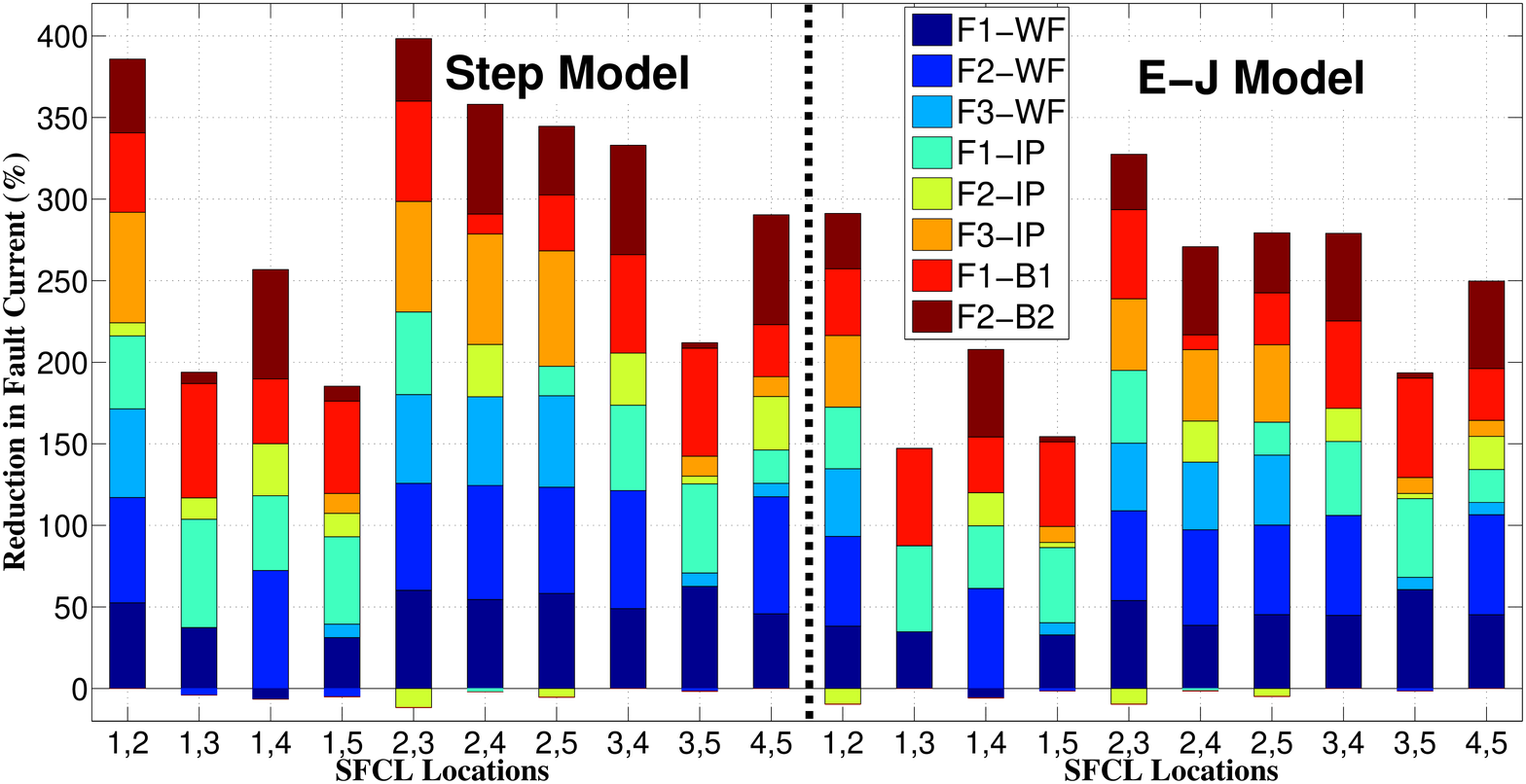}}
\caption{(Color Online) Reduction of the first peaks of fault currents achieved by different combinations of dual SFCLs.}
\label{Figure_8}
\end{center}
\end{figure*}
%%%%%%%%%%%%%%%%

On the other hand, for determining the optimal location where a single SFCL must be installed, the final decision has to be made under the circumstance of having a twofold conclusion. Firstly, the decision can be made accordingly to the highest total reduction on the fault current passing through different points and under different fault circumstances as shown in Fig.~\ref{Figure_7}. There, it can be observed that for the eight most important cases combining the occurence of a fault at certain positions and the measuring point for the current reduction, the SFCL installed at the port of the wind farm (Location 2) results to be the best option, as in this case the fault current can be reduced in six of the eight different scenarios with an accumulated reduction of 290\% from the step resistance model, and 220\% from the $E-J$ power law based model, respectively. Nevertheless, it is to be noticed that this strategy has also an adverse impact on the remaining two other scenarios (F1-IP \& F2-IP). Secondly, a decision can be made as well in terms of the overall performance for achieving positive impact under the scope of any of prospective circumstances.  Thus, we have determined that placing the SFCL at the Location 5, i.e, at the bus-tie between the industrial and domestic branches, is the most reliable option. In addition, by considering this strategy, the SFCL is capable of reducing the harmonics and voltage dips, doubling the short-circuit power, and ensuring even loading of parallel transformer. Furthermore, the recovery characteristics of the SFCL can also see benefit from this arrangement as after a quench of the SFCL, the bus-tie can be switched open for a short time (few seconds) to help the SFCL restoring the superconducting state. However, a drawback of the switching strategy is that this measure may temporarily reduce the quality of the power supply, but a strong impact on the normal operation of the power system is not foreseen.

\subsection{Dual SFCL's installation}~\label{SubSection_5_2}

For incorporating a dual strategy of protection by means of the installment of two SFCL, in Fig.~\ref{Figure_8} we present the current limiting performance of dual combinations of SFCLs allocated to different grid positions. According to both the step resistance model and the $E-J$ power law based model, the highest fault current reduction was always achieved when the SFCLs are installed at Location 2 (wind farm) and Location 3 (industrial branch) simultaneously, accomplishing a 400\% and 330\% total fault limitation respectively. Indeed, this arrangement can be considered as a much better strategy in comparison with the results obtained when just a single SFCL is considered, as the total current limitation is improved by $\sim110$\% and furthermore, contrary to the previous case, the current flowing through the integrating point when the fault occurs at the industrial branch (Fault 2) is significantly decreased instead of havind an adverse effect to the power system. Moreover, it is to be noticed that under this dual strategy, the current reduction measured shows a balanced performance on all the different analysed cases unlike the results obtained for when a sole SFCL is installed, what may also facilitate the designing of the control systems.

On the other hand, if the system operators measure the optimal strategy for the installation of the SFCLs in terms of the number of limited cases, different conclusions can be obtained under the framework of different physical models, i.e., when the step-resistance or $E-J$ power law based model is considered. Firstly, according to the step model, installing the SFCLs at Locations 1~\&~2 or Locations 4~\&~5 both can positively response to all eight measured fault conditions. In fact, when the SFCLs Locations 1~\&~2 are considered, a better performance is obtained as the total reduction in the fault current achieved under this arrangement is 40\% greater than the performance obtained when the SFCLs are installed at Locations  4~\&~5 (290\%). Secondly, when the $E-J$ power law model is assumed rather than the simplified step resistance model, placing the SFCLs at the Locations 1~\&~2 would increase the magnitude of current at the integrating point  under the occurrence of a fault in the domestic branch (Fault 2), it due to the unsuccessful triggering of the SFCL at Location 1 as explained in the previous subsection. Thus, from the point of view of the system operators the Locations 4~\&~5 can be considered as the most reliable solution as it is the only combination capable to limit all fault conditions and in all the considered scenarios.

\subsection{Cooperation of more than two SFCLs}~\label{SubSection_5_3}

%%%%%%%%%%%%%%%%
%%%%     FIGURE 9      %%%%%
%%%%%%%%%%%%%%%%
\begin{figure*}[t]
\begin{center}
{\includegraphics[width=1.0\textwidth]{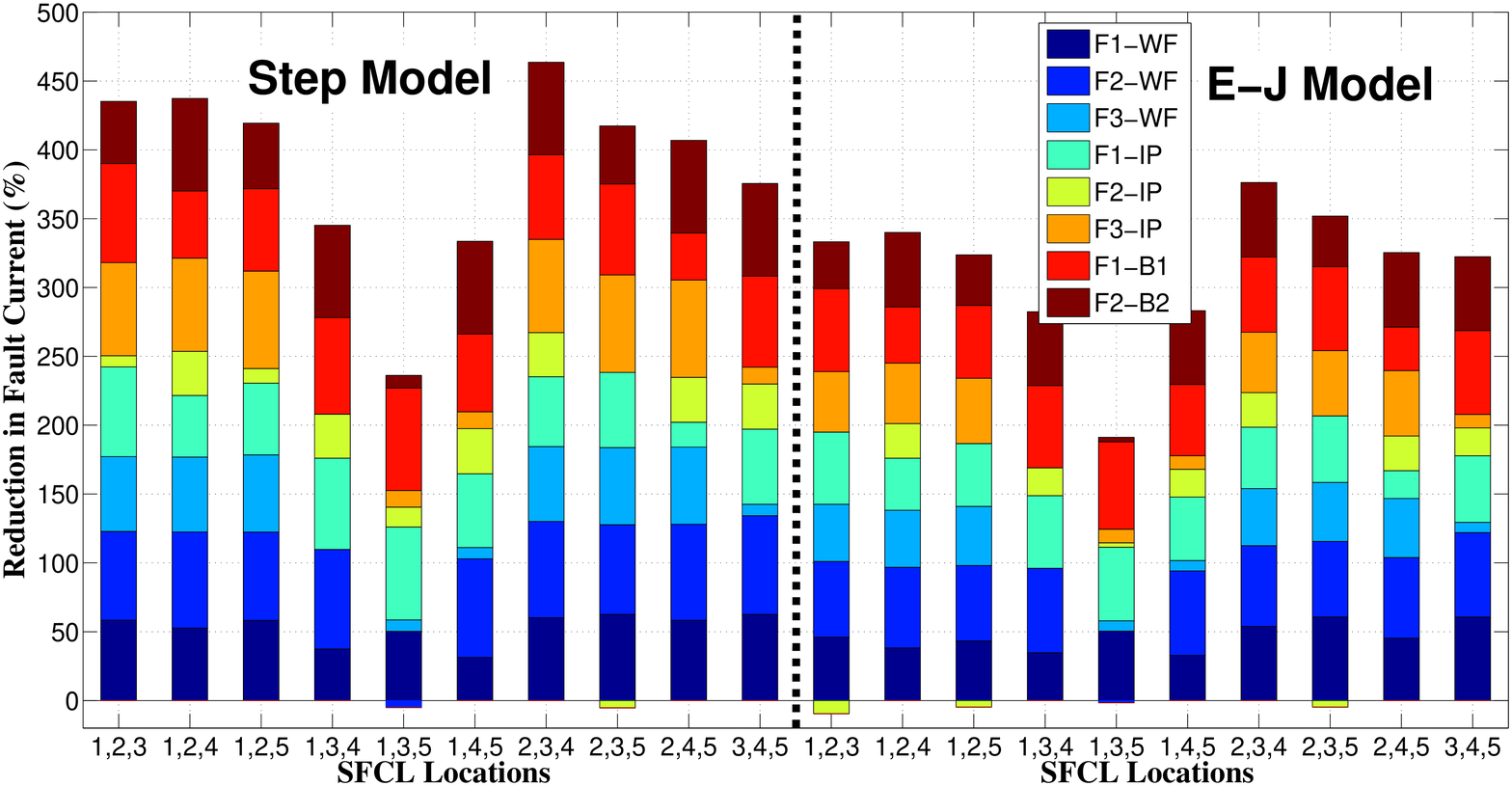}}
\caption{(Color Online) Total reduction of the first peaks of fault currents achieved by different combinations of three SFCLs.}
\label{Figure_9}
\end{center}
\end{figure*}
%%%%%%%%%%%%%%%%

As shown in Fig.~\ref{Figure_9}, most installation strategies of three SFCLs allow the reduction of the fault current at all eigth measured scenarios. Both SFCL models agree with the conclusion that the highest decrement in the fault current is achieved when the SFCLs are installed at the Locations 2, 3 and 4, simultaneously. This strategy shows a 470\% total reduction in the case of the step resistance model, and 375\% for the case of the $E-J$ power law based model, i.e., with attaining a significant increase on the overall performance of the system by about a 70\% and 45\% factor, respectively,  it in comparison with the best achieved performance when the dual SFCLs strategy is considered. Besides this remarkable improvement, the three SFCLs strategy can further response positively to any fault conditions, which means that for the concomitant decision between the current reduction criteria and the number of cases exhibiting fault current reduction, the choice for the three SFCLs strategy can be considered as the most reliable one. Nevertheless, until a significant reduction of the overall price of a SFCL will not be achieved, the distribution network operators could consider that this strategy may not be cost-effective in terms of the initial investment, but given the expected reduction on the prices of the second generation of high temperature superconducting wires, this decision can be seen as the most profitable strategy in terms of the grid safety and reliability. However, a limit for the maximum number of the SFCLs required must also be established in order to guarantee minimum costs with maximum benefits.

\begin{figure*}[t]
\begin{center}
{\includegraphics[width=1.0\textwidth]{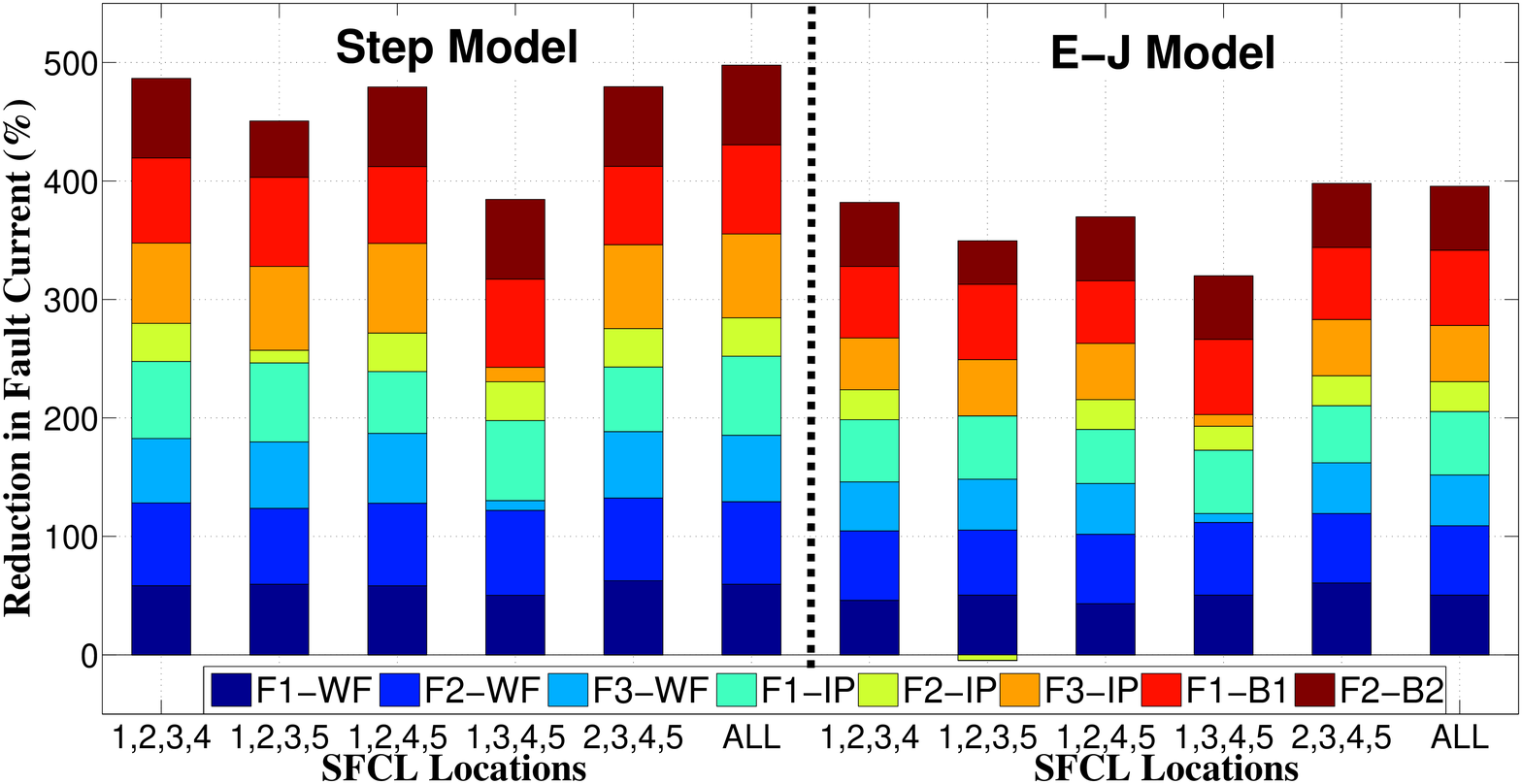}}
\caption{(Color Online) Total reduction of the first peaks of fault currents achieved by different combinations of four and five SFCLs.}
\label{Figure_10}
\end{center}
\end{figure*}

Fig.~\ref{Figure_10} shows the performance comparison among five different scenarios when four SFCLs are installed into the power system. With four SFCLs working together, all combinations can effectively limit the fault current for all eight studied cases excepting when the SFCLs are described by the $E-J$ power law based model and these are placed at the Locations 1, 2, 3, and 5, simultaneously. Under this scheme the fault current measured increases when the fault is initialized at the domestic branch (Fault 2), due to the no action of the SFCL installed at Location 1. On the other hand, it is to be noticed that when the step resistance model is considered, the accumulated maximum reduction on the fault current is again overestimated, achieving a 480\% reduction when the SFCLs are installed at the set of Locations $(1,2,3,4)$ or $(2,3,4,5)$, in comparison with a prospective reduction of 395\% at Locations $(2,3,4,5)$ when the most realist E-J power law based model is considered. In fact, even when an additional SFCL is installed, i.e., with the concurred action of up to 4 SFCLs, we have obtained that the accumulated maximum reduction on the fault current is just over a 10\% more than in the previous case (3 SFCLs), which allows to define an upper limit for the number of SFCLs needed as the 4 SFCLs alternative results not too convenient in terms of the added investment.

%%%%%%%%%%%%%%%%
%%%%     FIGURE 10    %%%%%
%%%%%%%%%%%%%%%%
%%%%%%%%%%%%%%%%

Nevertheless, in order to verify our previous statement, we have studied also the influence of considering even more SFCLs, as there is a total of five prospective locations for the SFCLs in the conventional power grid displayed in Fig.~\ref{Figure_1}. Compared with the last analyzed case (4 SFCLs),  the accumulated maximum reduction on the fault current has reached just a 15\% more increase when the SFCLs are based upon the step resistance model, but outstandingly no further improvement has been obtained when the more realistic power law based model with temperature dependence was incorporated.  This important result can be understood as a consequence of the mutual influence between the integrated SFCLs, i.e., when the fault current passing through one SFCL is substantially decreased by the influence of the others, the rate of heat accumulation slows down accordingly, leading to deceleration of the temperature rising and hence to a reduction of the resistance that the SFCL can develop before reaching the first peak of the fault.

Table~\ref{Table_1} summarizes the optimal allocation strategies and the corresponding performances of the SFCLs. The preferable locations for the installation of the SFCLs have been determined in terms of the two identified standards: $(i)$ the maximum accumulated fault current reduction and, $(ii)$ the maximum number of measuring conditions that could be limited. The results are categorized by the number of SFCLs that the system operators could want to install, and also the physical models that emulate the characteristics of the SFCLs. It is worth noting that in all the cases the step resistance model leads to an overestimation of the actual performance figures that may be offered by the SFCLs when more realistic physical properties are considered. Finally, we want to call the attention of the readers on the fact that when 
the strategy for maximizing benefits is installing a sole or two SFCLs, compromise has to be made between increasing the fault current reduction and the actual number of measuring conditions where the fault current can be limited. Therefore, based upon the comprehensive study presented in this article, we conclude that the optimal installation strategy refers to the installation of maximum 3 SFCLs at the Locations 2, 3, and 4, since for this case a maximum reduction of the fault current is achieved for all fault conditions and furthermore, adding more SFCLs does not represent a significant improvement concurrent with minimum investment requirements.

%%%%%%%%%%%%%%%%%%%%
%%%%%%        Table I          %%%%%%
%%%%%%%%%%%%%%%%%%%%

\begin{table*}%

\caption{\label{Table_1} Optimal installation strategies for SFCLs according to the step-resistance and E-J power law models. The maximum fault current reduction (FCR) value (per case) has been calculated as the sum of the percentage reductions of the fault current measured at the wind farm output, the integrated point, and branches 1 and 2, for the three fault conditions shown in Fig.~\ref{Figure_1}. It is worth noticing that, not at all measuring locations the fault current is reduced (see Figs.~\ref{Figure_7}--\ref{Figure_10}). Therefore, the table also shows the values for the accumulated fault current reduction when the fault current is reduced in a greater number of measuring conditions.}

%\raggedright

%\begin{flushleft}

\begin{tabularx}{\textwidth}{cccccc} \hline \hline

\textbf{Step-Resistance Model} \\

\hline \\

Maximum Fault Current Reduction (\%): & 290 & 400 & 470 & 480 & 495 \\

No. of measuring conditions with/without FCR: & 6 / 2 & 7 / 1 & 8 / 0 & 8 / 0 & 8 / 0 \\

Number of installed SFCLs: & 1 & 2 & 3 & 4 & 5 \\

SFCLs's Locations: & 2 & 2, 3 & 2, 3, 4 & 1, 2, 3, 4\footnotemark[1] & 1, 2, 3, 4, 5 \\

\\ \hline \\

No. of measuring conditions with/without FCR: & 7 / 1 & 8 / 0 & 8 / 0 & 8 / 0 & 8 / 0 \\

Accumulated FCR (\%) for Max. No. of measuring conditions: & 130 & 330 & 470 & 480 & 495 \\

Number of installed SFCLs: & 1 & 2 & 3 & 4 & 5 \\

SFCLs's Locations: & 5 & 1, 2 & 2, 3, 4 & 1, 2, 3, 4 & 1, 2, 3, 4, 5 \\

\\ \hline

\textbf{E-J Power Law Model} \\

\hline \\

Maximum Fault Current Reduction (\%): & 220 & 330 & 375 & 395 & 395 \\

No. of measuring conditions with/without FCR: & 6 / 2 & 7 / 1 & 8 / 0 & 8 / 0 & 8 / 0 \\

Number of installed SFCLs: & 1 & 2 & 3 & 4 & 5 \\

SFCLs's Locations: & 2 & 2, 3 & 2, 3, 4 & 2, 3, 4, 5 & 1, 2, 3, 4, 5 \\

\\ \hline \\

No. of measuring conditions with/without FCR: & 7 / 1 & 8 / 0 & 8 / 0 & 8 / 0 & 8 / 0 \\

Accumulated FCR (\%) for Max. No. of measuring conditions: & 120 & 250 & 375 & 395 & 395 \\

Number of installed SFCLs: & 1 & 2 & 3 & 4 & 5 \\

SFCLs's Locations: & 5 & 4, 5 & 2, 3, 4 & 2, 3, 4, 5 & 1, 2, 3, 4, 5 \\ 

\hline \hline

\end{tabularx}

%\end{flushleft}

\footnotetext[1]{Same performance is achieved when the four SFCLs are located at the positions 2, 3,4, and 5.}

\end{table*}

%%%%%%%%%%%%%%%%%%%%%%%%%%%%%%%%%%%%%%%
%%%%%%%%%%%% SECTION 6 %%%%%%%%%%%%%%%
%%%%%%%%%%%%%%%%%%%%%%%%%%%%%%%%%%%%%%% 

\section{Conclusion}~\label{Section_6}

The superconducting fault current limiter is a promising device to limit the escalating fault levels caused by the expansion of power grid and integration of renewables. This paper presents a comprehensive study on the performance and optimal allocation analysis of resistive type SFCLs inside of a power system with interconnected wind farm, built from the UK network standards. In order to unveil the impact of the superconducting material properties on the decision making for installing SFCLs, two different models have been considered throughout the study. Firstly, the active operation of the SFCL has been modeled by means of a step resistance or Heaviside function which is initialized by a set of preallocated parameters. Secondly, a most realistic model for the operation of the SFCL taking into consideration the proper $E-J$ characteristics of the superconducting material with dynamic temperature evolution has been considered. We have proven that SFCL technologies can effectively improve the damping characteristics of the generation system, and mitigate voltage dips at the grid, independently of the assumed model. However, we have demonstrated that despite a significant reduction on the time computing can be achieved when models of the step-resistance kind are considered, such simplifications lead to strong overestimations of the actual prospective performance of the SFCL, it in terms of the maximum reduction on the fault current and its correlated normal resistance. Furthermore, a complementary protection scheme for preventing the burning of the SFCL has been implemented together with the $E-J$ power law based model, what improves significantly the recovery of the SFCL during the transient states after a fault event.

Then, a systematic study on the prospective strategies for the installation of a sole or multiple SFCLs has been performed. Thence it has been proven that the concomitant cooperation of three SFCLs each installed at the Locations 2,  3, and 4, respectively, can be seen as the best protection strategy in terms of both the performance and the reliability figures of the overall grid within a minimum investment scheme vs maximum benefit. For achieving this conclusion all the potential combinations between two, three, four, and five SFCLs have been studied under a wide number of fault scenarios and measuring strategies.

%
%-------------------------%
%-------------------------%
%----- Bibliography ------%
%-------------------------%
%

\end{document}